\documentclass[aps,prd,twocolumn,showpacs,preprintnumbers,
               amsmath,amssymb]{revtex4}

\usepackage{graphicx}
\def\be{\begin{equation}}
\def\ee{\end{equation}}
\def\beqn{\begin{eqnarray}}
\def\eeqn{\end{eqnarray}}
\def\no{\nonumber}
\def\ba{\begin{array}{c}}
\def\bat{\begin{array}{cc}}
\def\ea{\end{array}}
\def\bi{\begin{itemize}}
\def\ei{\end{itemize}}

\def\cL{{\cal L}}

\def\cP{{\cal P}}

\def\cR{{\cal R}}

\newcommand{\e}{\mbox{\rm e}}

\begin{document}

\title{Yukawa Alignment in the Two-Higgs-Doublet Model}

\author{Antonio Pich} \author{Paula Tuz\'on}
\affiliation{Departament de F\'{\i}sica Te\`orica, IFIC,
Universitat de Val\`encia -- CSIC,
Apt. Correus 22085, E-46071 Val\`encia, Spain}

\begin{abstract}
\noindent
In multi-Higgs-doublet models the alignment in flavour space of the relevant Yukawa matrices
guarantees the absence of tree-level flavour-changing couplings of the neutral scalar fields.
We analyze the consequences of this condition within the two-Higgs-doublet model
and show that it leads to a generic Yukawa structure which contains as particular cases all known
specific implementations of the model based on ${\cal Z}_2$ symmetries.
All possible freedom in the Yukawa sector gets parametrized
in terms of three complex couplings $\varsigma_f$. In spite of having flavour conservation
in the neutral scalar couplings, the phases of these three parameters represent potential
new sources of CP violation.
\end{abstract}

\date{10 August 2009}


\maketitle


\section{Introduction}

The two-Higgs-doublet model (THDM) \cite{GHKD:90,BLS:99}
constitutes one of the simplest extensions of the Standard Model (SM) and incorporates
new and interesting phenomenological features. Many new-physics scenarios, including supersymmetry, can lead to
a low-energy spectrum containing the SM fields plus only one additional scalar doublet. Therefore, the THDM is also a very
convenient effective field theory framework to investigate low-energy effects of more generic dynamical settings.
The extended scalar sector accommodates two charged and three neutral scalar fields, in addition to the three Goldstones needed
to generate the gauge boson masses. This rich scalar spectrum provides a broad range of dynamical possibilities, such as potential
new sources of CP symmetry breaking, including spontaneous CP violation, axion phenomenology 
or dark matter candidates, just to mention a few.

In the most general version of the THDM, the fermionic couplings of the neutral scalars are non-diagonal in flavour.
The appearance of flavour-changing neutral current (FCNC) interactions represents a major shortcoming of the model.
Since FCNC phenomena are experimentally tightly constrained, one needs to implement ad-hoc dynamical restrictions
to guarantee the suppression of the FCNC couplings at the required level. For instance, the assumption that the non-diagonal
Yukawa couplings are proportional to the geometric mean of the two fermion masses, $g_{ij}\propto\sqrt{m_im_j}$ \cite{CS:87},
leads to a phenomenologically viable model usually known as `Type III' \cite{ARS:97}.
Models of this kind can be generated through particular textures of the Yukawa coupling matrices \cite{CS:87}.
Another obvious possibility is to require the scalar boson masses to be heavy enough to suppress any low-energy FCNC effects induced
through the scalar fields, but this drastically decreases the potential phenomenological relevance of the THDM.

A more elegant solution makes use of appropriately chosen discrete ${\cal Z}_2$ symmetries such that
only one scalar doublet couples to a given right-handed fermion field \cite{GW:77}. This guarantees the absence
of FCNCs in the Yukawa sector, but eliminates at the same time the possibility to have additional CP-violating phases beyond the
standard Kobayashi-Maskawa one. There are several possible models of this type, corresponding to different explicit implementations
of the ${\cal Z}_2$ symmetry. The minimal supersymmetric extension of the SM corresponds at tree level with one particular choice,
the so-called `Type II' THDM, which is the default version adopted in the majority of phenomenological analyses.

Flavour conservation in the neutral scalar couplings can be enforced in a rather trivial way requiring the Yukawa coupling matrices
to be aligned in flavour space. While this is also an ad-hoc constraint, lacking a proper dynamical explanation (probably coming from
a more fundamental high-energy theory), it leads to a much more general framework which contains as particular cases all known
THDMs based on ${\cal Z}_2$ symmetries. All possible freedom in the Yukawa sector gets parametrized in terms of three complex couplings
$\varsigma_f$;
their phases being possible new sources of CP violation. The general structure of this `aligned' THDM is discussed next, together with
some relevant phenomenological implications.

\section{The two-Higgs doublet model}
In its minimal version, that we are going to consider here, the THDM is an
$SU(3)_C\otimes SU(2)_L\otimes U(1)_Y$ theory with the fermion SM content (without right-handed neutrinos)
and two scalar doublets $\phi_a(x)$ ($a=1,2)$ with hypercharge $Y=\frac{1}{2}$.
The charge-conjugate fields $\tilde\phi_a(x) \equiv i \tau_2\,{\phi_a^*}$
are also $SU(2)$ doublets with $Y=-\frac{1}{2}$.
The neutral components of the two scalar doublets acquire vacuum expectation values
$\langle 0|\,\phi_a^T(x)|0\rangle = \frac{1}{\sqrt{2}}\, (0\, , v_a\,\e^{i\theta_a})$.
Without loss of generality, we can enforce $\theta_1=0$
through an appropriate $U(1)_Y$ transformation, leaving the relative phase $\theta\equiv\theta_2 - \theta_1$.
The gauge boson masses, which receive contributions from the two vacuum expectation values $v_a$,
are given by the same expressions than in the SM, with $v\equiv\sqrt{v_1^2+v_2^2}$.

Using a global $SU(2)$ transformation in the scalar space $(\phi_1,\phi_2)$,
it is possible to define a basis of scalar doublet fields (the so-called Higgs basis)
such that only $\Phi_1$ has non-zero vacuum expectation value ($\tan{\beta}\equiv v_2/v_1$):
\be
\left(\ba \Phi_1 \\ -\Phi_2\ea\right)\; \equiv\;
\left[\bat \cos{\beta} & \sin{\beta}\\ \sin{\beta} & -\cos{\beta}\ea\right]\;
\left(\ba\phi_1\\ \e^{-i\theta}\,\phi_2\ea\right)\, .
\ee
This has the advantage that the three Goldstone fields $G^\pm(x)$ and $G^0(x)$ get isolated as components of $\Phi_1$:
\be
\Phi_1 =\left[ \ba G^+\\ \frac{1}{\sqrt{2}}\left( v + S_1 + i G^0\right)\ea \right]\, ,
\quad
\Phi_2 =\left[ \ba H^+\\ \frac{1}{\sqrt{2}}\left( S_2 + i S_3\right)\ea \right]\, .
\no\ee
The physical scalar spectrum contains five degrees of freedom: the charged fields $H^\pm(x)$
and three neutral scalars $\varphi^0_i(x) = \left\{h(x), H(x), A(x)\right\}$, which are related through an orthogonal transformation
with the $S_i$ fields:
$\varphi^0_i(x) = \mathcal{R}_{ij} S_j(x)$.
The form of the $\mathcal{R}$ matrix is fixed by the scalar potential, which determines the neutral scalar mass matrix
and the corresponding mass eigenstates. In general, the CP-odd component $S_3$ mixes with the CP-even fields
$S_{1,2}$ and the resulting mass eigenstates do not have a definite CP quantum number.
If the scalar potential is CP symmetric this admixture disappears; in this particular case, $A(x) = S_3(x)$ and
\be
\left(\ba H\\ h\ea\right)\; = \;
\left[\bat \cos{(\alpha - \beta)} & \sin{(\alpha - \beta)} \\ -\sin{(\alpha - \beta)} & \cos{(\alpha - \beta)}\ea\right]\;
\left(\ba S_1\\ S_2\ea\right)\, .
\ee
\vskip .5cm

\section{Yukawa interactions}

The most general Yukawa Lagrangian is given by
\beqn\label{eq:GenYukawa}
\cL_Y & =&\mbox{}
 -\bar Q_L' (\Gamma_1\phi_1+\Gamma_2\phi_2)\, d_R'
-\bar Q_L' (\Delta_1\tilde\phi_1+\Delta_2\tilde\phi_2)\, u_R'
\no\\[5pt] &&\mbox{}
-\bar L_L' (\Pi_1\phi_1+\Pi_2\phi_2)\, l_R' + \mathrm{h.c.}\, ,
\eeqn
where $Q_L'$ and $L_L'$ denote the left-handed quark and lepton doublets. All fermionic fields are written as
$N_G$-dimensional flavour vectors;
i.e., $d_R' = (d_R', s_R', b_R',\cdots )$ and similarly for $u'_R$, $l_R'$, $Q_L'$ and $L_L'$.
The couplings $\Gamma_a$, $\Delta_a$ and $\Pi_a$ are
$N_G\times N_G$ complex matrices in flavour space.
In the Higgs basis, the Lagrangian takes the form
\beqn
\cL_Y & =& -\frac{\sqrt{2}}{v}\,\left\{
\bar Q_L' (M'_d\Phi_1+Y'_d\Phi_2) d_R'
\right.\no\\ &&\hskip .8cm\mbox{}
+\bar Q_L' (M'_u\tilde\Phi_1+Y'_u\tilde\Phi_2) u_R'
\no\\ &&\hskip .8cm\left.\mbox{}
+\bar L_L' (M'_l\Phi_1+Y'_l\Phi_2)\, l_R' \, +\, \mathrm{h.c.}\right\} ,
\eeqn
where $M'_f$ ($f=d,u,l$) are the non-diagonal fermion mass matrices, while the matrices $Y'_f$ contain
the Yukawa couplings to the scalar doublet with zero vacuum expectation value.

In the basis  
of fermion mass eigenstates $d(x)$, $u(x)$, $l(x)$, $\nu(x)$,
with diagonal mass matrices $M_f$ ($M_\nu=0$),
the corresponding matrices $Y_f$ are in general non-diagonal and unrelated to the fermion masses.
Thus, the Yukawa Lagrangian generates flavour-changing
interactions of the neutral scalars because there are two different Yukawa matrices
coupling to a given right-handed fermion field, which in general cannot be diagonalized simultaneously. The standard way
to avoid this problem is forcing one of the two matrices to be zero; i.e., imposing that only one scalar doublet
couples to a given right-handed fermion field \cite{GW:77}. This can be enforced implementing a discrete ${\cal Z}_2$ symmetry,
such that $\phi_1\to\phi_1$, $\phi_2\to -\phi_2$, $Q_L\to Q_L$, $L_L\to L_L$ and selecting appropriate
transformation properties for the right-handed fermion fields.
There are four non-equivalent possible choices, giving rise to the so-called
type-I (only $\phi_2$ couples to fermions) \cite{HKS:79,HW:81},
type-II ($\phi_1$ couples to $d$ and $l$, while $\phi_2$ couples to $u$) \cite{HW:81,DL:79},
leptophilic or type-X ($\phi_1$ couples to leptons and $\phi_2$ to quarks) and
type-Y ($\phi_1$ couples to $d$, while $\phi_2$ couples to $u$ and $l$) models \cite{BHP:90,GR:94,AS:95,AKTY:09}.
The explicit implementation of the ${\cal Z}_2$ symmetry is scalar-basis dependent. If
the ${\cal Z}_2$ symmetry is imposed in the Higgs basis, all fermions are forced to couple to the field
$\Phi_1$ in order to get non-vanishing masses.
This {\it inert} 
doublet model provides a natural frame for dark matter
\cite{MA:08,BHR:06,LNOT:07}; note however that
although $\Phi_2$ does not couple to fermions, it does have electroweak interactions.

A softer and more general way to avoid tree-level FCNC interactions is to require the alignment in flavour space of the
Yukawa couplings of the two scalar doublets.  It is convenient to implement this condition in the form:
\be
\Gamma_2 =\xi_d\,\e^{-i\theta} \,\Gamma_1\, ,
\quad
\Delta_2 =\xi_u^*\,\e^{i\theta} \,\Delta_1\, ,
\quad
\Pi_2 =\xi_l\,\e^{-i\theta} \,\Pi_1\, .
\ee
The proportionality parameters $\xi_f$ are arbitrary complex numbers. To simplify later equations, we have redefined these
parameters introducing the explicit phases $\e^{\mp i\theta}$ which cancel the relative global phases between the two scalar doublets.
The Yukawa alignment guarantees that the $Y'_f$ and $M'_f$ matrices are proportional and, therefore, can be simultaneously
diagonalized with the result:
\be
Y_{d,l} = \varsigma_{d,l}\, M_{d,l}\, ,
\quad
Y_{u} = \varsigma_{u}^*\, M_{u}\, ,
\quad
\varsigma_f\equiv \frac{\xi_f - \tan{\beta}}{1+\xi_f\tan{\beta}}\, .
\ee

In terms of the mass-eigenstate fields, the Yukawa interactions take then the form:
\beqn\label{eq:alignedY}
\cL_Y &\!\! =&\!\!
-\frac{\sqrt{2}}{v}\; H^+\! (x) \,\bar u(x) \left[\varsigma_{d}\, V M_{d} \cP_R -
\varsigma_{u}\, M_{u} V \cP_L\right] d(x)
\no\\ &&\!\! 
-\frac{\sqrt{2}}{v}\; H^+\! (x) \;\varsigma_{l}\;\bar\nu(x)  M_l\,\cP_R\, l(x)
\no\\[7pt] &&\!\! 
-\frac{1}{v}\; \sum_{\varphi^0_i, f} 
\varphi^0_i(x)\; 
y_f^{\varphi^0_i}\; \bar f(x)\, M_f\, \cP_R\,  f(x)
\, +\,\mathrm{h.c.}
\eeqn
where $V$ is the Cabibbo-Kobayashi-Maskawa quark mixing matrix and $\cP_{R,L}\equiv\frac{1}{2}(1\pm\gamma_5)$
the chirality projectors.

The flavour alignment of the Yukawa couplings results in a very specific structure for the scalar-fermion interactions:
\begin{enumerate}

\item[i)] All fermionic couplings of the physical scalar fields are proportional to the corresponding fermion mass matrices.

\item[ii)] The neutral Yukawas are diagonal in flavour. The couplings of the physical scalar fields $H$, $h$ and $A$ are obviously
proportional to the corresponding elements of the orthogonal matrix $\mathcal{R}$,
\beqn
y_{d,l}^{\varphi^0_i} &=& \cR_{i1} + (\cR_{i2} + i\,\cR_{i3})\,\varsigma_{d,l}  \, ,
\no\\ 
y_u^{\varphi^0_i} &=& \cR_{i1} + (\cR_{i2} -i\,\cR_{i3}) \,\varsigma_{u}^*  \, .
\eeqn

\item[iii)] The only source of flavour-changing phenomena is the
quark-mixing matrix $V$, which regulates the quark couplings of the $W^\pm$ gauge bosons and the charged scalars $H^\pm$.

\item[iv)] All leptonic couplings are diagonal in flavour. This is obviously related to the absence of right-handed neutrino fields in our low-energy Lagrangian.
    Since neutrinos are massless, the leptonic mixing matrix $V_L$ can be reabsorbed through a redefinition of the neutrino fields: \
    $\bar\nu \cdot V_L\to\bar\nu$.

\item[v)] The only new couplings introduced by the Yukawa Lagrangian are the three parameters $\varsigma_f$, which encode all possible freedom allowed by the alignment conditions. These couplings satisfy universality among the different generations: all fermions of a given electric charge have the same universal coupling $\varsigma_f$. Moreover, the parameters $\varsigma_f$ are invariant under global SU(2) transformations of the scalar fields,
$\phi_a\to \phi_a' = U_{ab}\phi_b$ \cite{DH:05}; i.e., they are independent of the basis choice adopted in the scalar space.

\item[vi)] The usual models with a single scalar doublet coupling to each type of right-handed fermions are recovered taking the appropriate limits $\xi_f\to 0$ or $\xi_f\to\infty$ ($1/\xi_f\to 0$); i.e.,
$\varsigma_f \to -\tan{\beta}$ or  $\varsigma_f \to \cot{\beta}$.
The type-I model corresponds to $(\xi_d,\xi_u,\xi_l)=(\infty,\infty,\infty)$,
type II to $(0,\infty,0)$, type X to $(\infty,\infty,0)$ and type Y to $(0,\infty,\infty)$.
The {\it inert} doublet model
corresponds to
$\varsigma_f=0$ ($\xi_f = \tan{\beta}$).
The $\varsigma_f$ values for all these particular models based on ${\cal Z}_2$
symmetries are given in Table~\ref{tab:z2models}.

\item[vii)] The $\varsigma_f$ can be arbitrary complex numbers, opening the possibility to have new sources
of CP violation without tree-level FCNCs.

\end{enumerate}

The Yukawa alignment provides a general setting to discuss the phenomenology of THDMs without tree-level FCNCs, parameterizing
the different possibilities through the three complex couplings $\varsigma_f$.
\vskip .5cm

\begin{table}[tb]\centering
\begin{tabular}{|c|c|c|c|}
\hline
Model & $\varsigma_d$ & $\varsigma_u$ & $\varsigma_l$
\\ \hline
Type I & $\cot{\beta}$ & $\cot{\beta}$ & $\cot{\beta}$
\\
Type II & $-\tan{\beta}$ & $\cot{\beta}$ & $-\tan{\beta}$
\\
Type X & $\cot{\beta}$ & $\cot{\beta}$ & $-\tan{\beta}$
\\
Type Y & $-\tan{\beta}$ & $\cot{\beta}$ & $\cot{\beta}$
\\
Inert & 0 & 0 & 0
\\ \hline
\end{tabular}
\caption{Choices of couplings $\varsigma_f$ which correspond to models with
discrete ${\cal Z}_2$ symmetries.}\label{tab:z2models}
\end{table}

\section{Quantum corrections}

Quantum corrections could introduce some misalignment of the Yukawa coupling matrices, generating small FCNC effects suppressed
by the corresponding loop factors. However, the special structure of the aligned THDM strongly constrains the possible FCNC
interactions.

The Lagrangian of the aligned THDM is invariant under flavour-dependent phase transformations of the fermion
mass eigenstates,
$f_i(x)\to\e^{i\alpha^f_i}\, f_i(x)$,
provided the quark mixing matrix is transformed as
$V_{ij}\to\e^{i\alpha^u_i} V_{ij}\, \e^{-i\alpha^d_j}$.
Here $f=d,u,l,\nu$, while the subindex $i$ refers to the different fermion generations.
Since the equivalent lepton mixing matrix has been reabsorbed into the neutrino fields, the redefined $\nu_i(x)$ fields
have $\alpha^\nu_i = \alpha^l_i$.
Owing to this symmetry, lepton-flavour-violating neutral couplings are identically zero to all orders in perturbation theory, because
there is neither lepton mixing in the charged-current couplings in the absence of right-handed neutrinos.
The usually adopted ${\cal Z}_2$ symmetries are unnecessary in the lepton sector, making the model variants `Type X'
and `Type Y' less compelling. From a phenomenological point of view, the coupling $\varsigma_l$ can take any value.
In fact, a leptophobic (or lepton-inert) model with $\varsigma_l=0$ is a perfectly acceptable possibility, independently
of the assumed structure of the quark couplings. Thus, one could easily evade all phenomenological constraints coming from leptonic or
semileptonic processes.

In the quark sector the quark-mixing matrix can induce FCNC structures of the type $\bar u_i F^u_{ij} u_j$
or $\bar d_i F^d_{ij} d_j$. To preserve the symmetry under phase redefinitions of the quark fields, the matrices $F^q_{ij}$ should
transform as
$F^u_{ij}\to\e^{i\alpha^u_i} F^u_{ij}\, \e^{-i\alpha^u_j}$ \ and \
$F^d_{ij}\to\e^{i\alpha^d_i} F^d_{ij}\, \e^{-i\alpha^d_j}$.
This determines $F^q_{ij}$ to adopt the forms
$F^u_{ij} = \sum_{k} V_{ik}\, g(m_{d_k}) V^\dagger_{kj}$
and
$F^d_{ij} = \sum_{k} V^\dagger_{ik}\, \tilde g(m_{u_k}) V_{kj}$,
or similar ones with more factors of $V$ and $V^\dagger$.
For $i\not= j$, the unitarity of $V$ (GIM mechanism) makes necessary the presence of mass-dependent factors
through the loop functions $g(m_{d_k})$ and $\tilde g(m_{u_k})$.
The only possible local FCNC terms are then of the type ($n,m>0$)
$F^u = V (M_d)^n V^\dagger$ and
$F^d = V^\dagger (M_u)^m V$
(or similar structures with additional factors of $V$ and $V^\dagger$).
Therefore, the flavour mixing induced by loop corrections has a very characteristic structure which resembles
the popular Minimal Flavour Violation scenarios \cite{MFV}.
Notice however that local FCNC terms shall disappear
whenever the $\varsigma_f$ parameters approach any of the different sets of values given in Table~\ref{tab:z2models},
because these limits are protected by corresponding ${\cal Z}_2$ symmetries (provided the scalar potential
is ${\cal Z}_2$ symmetric). One-loop FCNC effects are then very constrained.
\vskip .5cm

\section{Discussion} 

One of the most distinctive features of the THDM is the presence of a charged scalar.
The process $e^+e^-\to H^+H^-$ provides a very
efficient way to search for $H^\pm$  
without referring to any specific Yukawa structure.
Assuming that $\mathrm{Br}(H^+\to c\bar s)+ \mathrm{Br}(H^+\to \tau^+\nu_\tau)=1$,
the combined LEP data constrain $M_{H^\pm} > 78.6$~GeV (95\% CL) \cite{LEP:01}.
The limit slightly improves if particular values of $\mathrm{B}_\tau\equiv\mathrm{Br}(H^+\to \tau^+\nu_\tau)$ are assumed.
The weakest limit of 78.6~GeV is obtained for $\mathrm{B}_\tau\sim 0.4$--$0.5$, improving to 81.0 (89.6) for $\mathrm{B}_\tau=0$ (1)
which corresponds to $\varsigma_l=0$ ($\varsigma_{u,d}=0$).
These limits could be avoided for a fermiophobic (inert) THDM with $\varsigma_f \ll 1$. The charged scalar could then be detected through the decay mode $H^\pm\to W^\pm A$, provided it is kinematically allowed. Assuming a CP-conserving scalar potential,
OPAL finds the (95\% CL) constraints
$M_{H^\pm} > 56.5\; (64.8)$~GeV  for $12\; (15)~\mathrm{GeV} < M_A < M_{H^\pm} - M_{W^\pm}$ \cite{OPAL:08}.

CDF \cite{CDF:09} and D0 \cite{D0:09} have searched for $t\to H^+ b$ decays with negative results.
CDF assumes $\mathrm{Br}(H^+\to c\bar s)=1$ ($\varsigma_l\ll\varsigma_{u,d}$), while D0 adopts the
opposite hypothesis $\mathrm{Br}(H^+\to \tau^+\nu_\tau)=1$ ($\varsigma_l\gg\varsigma_{u,d}$). Both experiments
find upper bounds on $\mathrm{Br}(t\to H^+ b)$ around 0.2 (95\% CL) for charged scalar masses between 60 to 155 GeV.
This implies $|\varsigma_u| < 0.3$--2.5, the exact number depending on the particular value of $M_{H^\pm}$ within the
analyzed range.

Weak decays provide valuable probes of virtual $H^\pm$ contributions.
The $\tau\to\mu/e$ lepton universality test
$|g_\mu/g_e| = 1.0000\pm 0.0020$ \cite{PI:08} gives the lower bound
$M_{H^\pm}/|\varsigma_l| >  1.78~\mathrm{GeV}$  (90\% CL).
Owing to the helicity suppression of the SM amplitude, the  semileptonic decays $P^+\to l^+\nu_l$ are more sensitive to $H^+$ exchange:
$ \Gamma(P_{ij}^+\to l^+\nu_l)/\Gamma(P_{ij}^+\to l^+\nu_l)_{\mathrm{SM}} = \left|1-\Delta_{ij} \right|^2$,
with $\Delta_{ij} = (m_{P_{ij}^\pm}/M_{H^\pm})^2 \varsigma^*_l (\varsigma_u m_{u_i}+\varsigma_d m_{d_j})/(m_{u_i}+m_{d_j})$.
The correction $\Delta_{ij}$ is in general complex and its real part could have either sign.
To determine its size one needs to know $V_{ij}$
(extracted from $P_{l3}$ decays which are much less sensitive to scalar contributions)
and a theoretical determination of the meson decay constant. From the present knowledge on $B^+\to\tau^+\nu_\tau$
\cite{CKM:08}, we obtain
$|1-\Delta_{ub}| =1.32 \pm 0.20$ which implies $M_{H^\pm}/\sqrt{|\mathrm{Re}(\varsigma_l^*\varsigma_d)|} > 5.7$~GeV  (90\% CL).
Using $f_{D_s} = 242\pm 6$~MeV \cite{f_Ds}, the
recent CLEO measurement $\mathrm{Br}(D_s^+\to\tau^+\nu_\tau) = (5.62\pm 0.44)\%$ \cite{CLEO:09} implies
$|1-\Delta_{cs}| =1.07 \pm 0.05$ and $M_{H^\pm}/\sqrt{|\mathrm{Re}(\varsigma_l^*\varsigma_u)|} > 4.4$~GeV  (90\% CL).

FCNC processes induced at the quantum level, such as $b\to s\gamma$ or $P^0$--$\bar P^0$ mixing, are also very sensitive
to scalar contributions and provide strong constraints on the parameter space ($M_{H^\pm},\varsigma_u,\varsigma_d$).
In the CP-conserving limit, these type of transitions have been already analyzed within THDMs with discrete
$\mathcal{Z}_2$ symmetries \cite{AKTY:09,Mahmoudi:2009zx} or adopting the `Type III' ansatz \cite{ARS:97,Mahmoudi:2009zx}.
A detailed phenomenological analysis within the more general aligned THDM is underway.
Of particular interest are the CP-violating effects induced by the complex phases of the $\varsigma_f$
parameters. They represent a new source of CP violation in the Yukawa sector which does not introduce
any dangerous FCNC couplings at tree level.
The structure of the aligned THDM at higher orders is also worth investigating. While loop corrections could generate
some misalignment of the Yukawa structures, the resulting FCNC phenomena are strongly constrained by the
symmetries of the aligned THDM Lagrangian. We plan to analyze these questions in future publications.

%
\section*{Acknowledgements}
This work has been supported in part by
the EU Contract MRTN-CT-2006-035482 (FLAVIAnet),
by MICINN, Spain [grants FPA2007-60323 and Consolider-Ingenio 2010 CSD2007-00042 –CPAN–]
and by Generalitat Valenciana (PROMETEO/2008/069). P.T is indebted to the Spanish MICINN for a FPU
fellowship.



\begin{thebibliography}{10}

\bibitem{GHKD:90} J. Gunion, H.E. Haber, G. Kane, and S. Dawson, {\it The
Higgs Hunter's Guide} (Addison-Wesley, New York, 1990).

\bibitem{BLS:99} G.C. Branco, L. Lavoura, and J.P. Silva, {\it CP Violation}
(Oxford University Press, Oxford, England, 1999).

\bibitem{CS:87} T.P. Cheng and M. Sher,  {\it Phys. Rev.} {\bf D35}, 3484 (1987).
J.L. Diaz-Cruz, R. Noriega-Papaqui and A. Rosado,
{\it Phys. Rev.} {\bf D69}, 095002 (2004).


\bibitem{ARS:97} D. Atwood, L. Reina and A. Soni,
  {\it Phys. Rev.} {\bf D55}, 3156 (1997).

\bibitem{GW:77} S.L. Glashow and S. Weinberg, {\it Phys. Rev.} {\bf D15}, 1958 (1977).

\bibitem{HKS:79} H.E. Haber, G.L. Kane and T. Sterling,
   {\it Nucl. Phys.} {\bf B161}, 493 (1979).

\bibitem{HW:81} L.J. Hall and M.B. Wise, {\it Nucl. Phys.} {\bf B187}, 397 (1981).

\bibitem{DL:79} J.F. Donoghue and L.F. Li, {\it Phys. Rev.} {\bf D19}, 945 (1979).

\bibitem{BHP:90} V.D. Barger, J.L. Hewett and R.J.N. Phillips,
   {\it Phys. Rev.} {\bf D41}, 3421 (1990).

\bibitem{GR:94} Y. Grossman, {\it Nucl. Phys.} {\bf B426}, 355 (1994).

\bibitem{AS:95} A.G. Akeroyd and W.J. Stirling, {\it Nucl. Phys.} {\bf B447}, 3 (1995).
A.G. Akeroyd, {\it Phys. Lett.} {\bf B377} (1996) 95; {\it J.~Phys.} {\bf G24}, 1983 (1998).

\bibitem{AKTY:09} M. Aoki, S. Kanemura, K. Tsumura and K. Yagyu,
{\it Phys. Rev.} {\bf D80}, 015017 (2009).  

\bibitem{MA:08} E. Ma, {\it Mod. Phys. Lett.} {\bf A23}, 647 (2008). 
{\it Phys. Rev.} {\bf D73}, 077301 (2006).
N.G. Deshpande and E. Ma, {\it Phys. Rev.} {\bf D18}, 2574 (1978).

\bibitem{BHR:06} R. Barbieri, L.J. Hall and V.S. Rychkov,
  {\it Phys. Rev.} {\bf D74}, 015007 (2006).

\bibitem{LNOT:07} L. L\'opez Honorez, E. Nezri, J.F. Oliver and M.H.G. Tytgat,
{\it JCAP} {\bf 0702}, 028 (2007).

\bibitem{DH:05} S. Davidson and H.E. Haber, {\it Phys. Rev.} {\bf D72}, 035004 (2005); Erratum-ibid. {\bf D72}, 099902 (2005). 
I.F. Ginzburg and M. Krawczyk, {\it Phys. Rev.} {\bf D72}, 115013 (2005).
F.J.~Botella and J.P. Silva, {\it Phys. Rev.} {\bf D51}, 3870 (1995).

\bibitem{MFV}
R.S.~Chivukula and H.~Georgi, {\it Phys. Lett.} {\bf B188}, 99 (1987).
L.J.~Hall and L.~Randall, {\it Phys. Rev. Lett.} {\bf 65}, 2939 (1990).
A.J.~Buras et al., {\it Phys. Lett.} {\bf B500}, 161 (2001).
%
  G.~D'Ambrosio, G.~F.~Giudice, G.~Isidori and A.~Strumia,
  {\it Nucl. Phys.} {\bf B645}, 155 (2002).
%
  V.~Cirigliano, B.~Grinstein, G.~Isidori and M.~B.~Wise,
  {\it Nucl. Phys.} {\bf B728}, 121 (2005).


\bibitem{LEP:01} ALEPH, DELPHI, L3 and OPAL Collaborations, The LEP Working Group for Higgs Boson Searches,
LHWG note 2001-05 [arXiv:hep-ex/0107031].

\bibitem{OPAL:08} OPAL Collaboration, arXiv:0812.0267 [hep-ex].

\bibitem{CDF:09} CDF Collaboration, {\it Phys. Rev. Lett.} {\bf 103}, 101803 (2009).

\bibitem{D0:09} D0 Collaboration, {\it Phys. Rev.} {\bf D80}, 051107 (2009).

\bibitem{PI:08} A. Pich, {\it Nucl. Phys. B (Proc. Suppl.)} {\bf 181-182}, 300 (2008).

\bibitem{CKM:08} M. Antonelli et al., arXiv:0907.5386 [hep-ph].

\bibitem{f_Ds}
  E.~Follana, C.T.H.~Davies, G.P.~Lepage and J.~Shigemitsu
  {\it Phys. Rev. Lett.} {\bf 100}, 062002 (2008).
  C. Bernard et al., PoS LATTICE2008, 278 (2008).
  B.~Blossier et al., {\it JHEP} {\bf 0907}, 043 (2009).  

\bibitem{CLEO:09} CLEO Collaboration, {\it Phys. Rev.} {\bf D79}, 052001 (2009); ibid. {\bf D79}, 052002 (2009). 

\bibitem{Mahmoudi:2009zx} F.~Mahmoudi and O.~Stal,
  arXiv:0907.1791 [hep-ph].


\end{thebibliography}
\end{document}